\documentclass[conference]{IEEEtran}
\IEEEoverridecommandlockouts
\usepackage{cite}
\usepackage{amsmath,amssymb,amsfonts}
\usepackage{algorithm}
\usepackage{algpseudocode}
\usepackage{graphicx}
\usepackage{textcomp}
\usepackage{xcolor}
\usepackage{subcaption}
\usepackage{hyperref}
\def\BibTeX{{\rm B\kern-.05em{\sc i\kern-.025em b}\kern-.08em
    T\kern-.1667em\lower.7ex\hbox{E}\kern-.125emX}}

\begin{document}


\title{Self-supervised Radio Representation Learning: \\Can we Learn Multiple Tasks?}





\author{\IEEEauthorblockN{ Ogechukwu Kanu\IEEEauthorrefmark{3}, Ashkan Eshaghbeigi\IEEEauthorrefmark{1}, and
Hatem Abou-Zeid\IEEEauthorrefmark{3}
}
\IEEEauthorblockA{\IEEEauthorrefmark{3}{Department of Electrical and Software Engineering}, 
{University of Calgary}, Canada}
\IEEEauthorblockA{\IEEEauthorrefmark{1}{Qoherent Inc.}, 
{Toronto, Ontario, Canada}
\thanks{This research is supported by Alberta Innovates, MITACS Accelerate, and Qoherent Inc.
The authors also thank Denvr Dataworks, Calgary, Canada, for their high-performance compute.
}}

}

\maketitle

\begin{abstract}

Artificial intelligence (AI) is anticipated to play a pivotal role in 6G. 
However, a key challenge in developing  AI-powered solutions is the extensive data collection and labeling efforts required to train supervised deep
learning models. To overcome this, self-supervised learning (SSL) approaches have recently demonstrated remarkable success across various domains by 
leveraging large volumes of unlabeled data to achieve near-supervised performance. 
In this paper, we propose 
an effective SSL scheme for radio signal representation learning using momentum contrast. By applying contrastive learning, our method extracts robust, transferable representations from a large real-world dataset. We assess the generalizability of these learned representations across two wireless communications tasks: angle of arrival (AoA) estimation and automatic modulation classification (AMC). 
Our results show that carefully designed augmentations and diverse data enable contrastive learning to produce high-quality, invariant latent representations.
These representations are effective even with frozen encoder weights, and fine-tuning further enhances performance, surpassing supervised baselines.
To the best of our knowledge, this is the first work to propose and 
demonstrate the effectiveness of
self-supervised learning for radio signals across multiple
tasks. 
Our findings highlight the potential of self-supervised learning to transform AI for wireless communications by reducing dependence on labeled data and improving model generalization $-$ paving the way for scalable \emph{foundational 6G AI models} and solutions. 


\end{abstract}

\begin{IEEEkeywords}
Self-supervised learning, 6G, foundational models, artificial intelligence, direction of arrival, contrastive learning.
\end{IEEEkeywords}

\section{Introduction}

Following the widespread deployment and success of the fifth generation (5G) networks, attention has shifted towards the sixth-generation (6G) wireless communication systems \cite{ cui2024overview}. It is anticipated that 6G will meet the ultra-low latency, higher data rates, ultra-reliable and ubiquitous connectivity demands in wireless communication systems. In parallel, AI technology has advanced rapidly over the years, particularly with successes in natural language processing (NLP), vision, robotics, and foundation models. 

Several of these AI advances have no doubt found their way to wireless communication networks. In the 5G era, AI has been successfully applied to many use-cases including network optimization, traffic prediction, fault detection, significantly enhancing network performance and user experience \cite{cui2024overview}. 
It is expected that AI will further redefine wireless communication in 6G, transitioning networks from mere connectivity enablers to intelligent ecosystems.

However, conventional AI-driven models, predominantly based on supervised learning, have demonstrated limited scalability and adaptability in high stakes applications where labelled data is scarce or expensive to obtain. Self-supervised learning (SSL) has emerged as a powerful alternative that harnesses large volumes of unlabelled data and thus reduces the dependency on manual annotations. It has enabled foundation models across domains, empowering AI models to learn meaningful representations without explicit labels - which subsequently enhances generalization and robustness. Foundation models in vision have applied self-supervised learning techniques without need for labels and addressed tasks like segmentation, classification and object detection in zero-shot or few-shot learning contexts \cite{wang2024sam}. In multimodal learning, techniques like CLIP and SAM illustrate the versatility of self-supervised foundation models that adapt seamlessly across various tasks without retraining \cite{zong2023self}.

Given the abundance of unlabelled radio data, adopting self-supervised learning in wireless communications is well motivated and requires further research. Some recent studies have explored its application in various wireless localization and sensing tasks  
\cite{alloulah2022self, roldan2023self, davaslioglu2022self, li2022unsupervised, hao2024bootstrapping}, channel estimation \cite{zhang2023self}
, and RF fingerprinting \cite{shao2024specific}.
However, the primary focus of prior work in wireless communications has been on a single task. In this paper, we ask the question of whether SSL can be used effectively for multiple tasks in the wireless context. 
We draw inspiration from the recent advances in representation learning from other domains and propose the usage of momentum contrast for \emph{generic} radio representation learning. Our goal is two-fold: 1) to devise a robust scheme of SSL using momentum contrast to learn radio representations with radio-specific augmentations, and 2) to assess the transferability of the learned representations across multiple wireless tasks using real-world datasets.
The following are the contributions of this paper:
\begin{itemize}
    \item To the best of our knowledge, this work is the first to propose and demonstrate the effectiveness of self-supervised learning on radio signals for multiple tasks. We achieve a robust performance across two downstream tasks: angle of arrival (AoA) estimation and automatic modulation classification (AMC) on a real-world dataset. Our results show that our proposed framework learns radio representations that are effective even with frozen encoder weights, and fine-tuning further enhances performance, surpassing supervised baselines.
    \item We demonstrate the strength of the proposed SSL framework for low-labeled data regimes, achieving up to 41\% improvement over fully supervised baselines and exceeding performance with the full scale of data. We also conduct a thorough evaluation of the impact of model capacity and the number of classifier layers on the performance of the downstream task. 
    \item This work is the first to introduce antenna dropout as an augmentation technique in wireless communications operating directly on received IQ data streams.  This is inspired by it’s effectiveness in SSL for radar-sensing demonstrated in \cite{hao2024bootstrapping}. 
    
    \item This work also introduces a significant contribution through the evaluation on real-world datasets, gathered using a purpose-built testbed. 70 GBytes of IQ wireless data were captured encompassing a range of modulation types. 
\end{itemize}

The paper is structured as follows. Section II presents related work while Section III introduces the problem statement and the momentum contrastive learning methodology. We then detail the experiment setup, the dataset, and performance evaluation in Section IV. 
The results are presented and discussed in Section V and we end with conclusions in Section VI.

\section{Related Work}
Recently, a few studies have investigated the use of self-supervised learning for wireless signals. In \cite{alloulah2022self}, a co-learning scheme combining radar and vision uses contrastive learning to train a radar-only sensing model, showing comparable performance to supervised models with full labels and achieving a 3.5\% gain using just 1\% of labels, with further gains anticipated for complex tasks. Similarly, \cite{li2022unsupervised} applied contrastive and predictive learning to human sensing tasks (e.g., pose estimation, re-identification, and action recognition) using RF signals as 2D heatmaps, reporting gains up to 5.7\% over supervised learning for the same amount of labels and 11.3\% with extra unlabelled data. They found that predictive learning offers better transferability than contrastive learning.
Other studies, such as \cite{davaslioglu2022self},
show success using contrastive learning on wireless received data, which similarly, enhances data efficiency with fewer labels compared to supervised methods.

The study in \cite{roldan2023self} employs a reconstructive self-supervised learning approach to improve angular resolution without modifying the physical hardware. They train a neural network to extrapolate the responses of additional antenna elements based on existing ones, effectively enlarging the antenna aperture. These studies show the benefits of using self-supervised learning and highlight some challenges, particularly with contrastive learning. 
Unlike existing works and to the best of our knowledge, we are the first to present the effectiveness of self-supervised learning for radio representation learning towards multitask foundation radio models. We demonstrate that indeed even when the SSL encoder weights are frozen very good performance can be achieved - and subsequent fine-tuning achieves competitive performance compared to supervised solutions with a fraction of the data requirements. 

\section{Self-Supervised Radio Representation Learning with Momentum Contrast}

\begin{figure}[!b]
     \centering
     \includegraphics[width=0.75\linewidth]{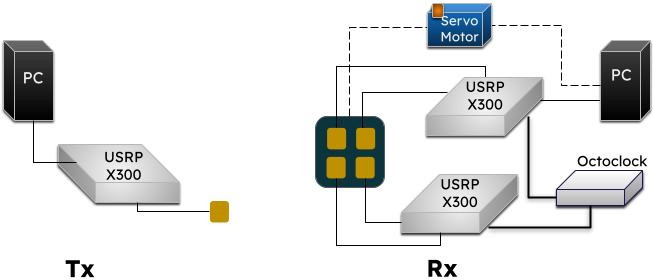}
     \caption{MIMO testbed used to create the dataset used through our study.}
     \label{fig:testbed}
 \end{figure}

\begin{figure*}[!t]
    \begin{subfigure}[b]{0.55\linewidth}  
        \centering
        \includegraphics[width=1\linewidth]{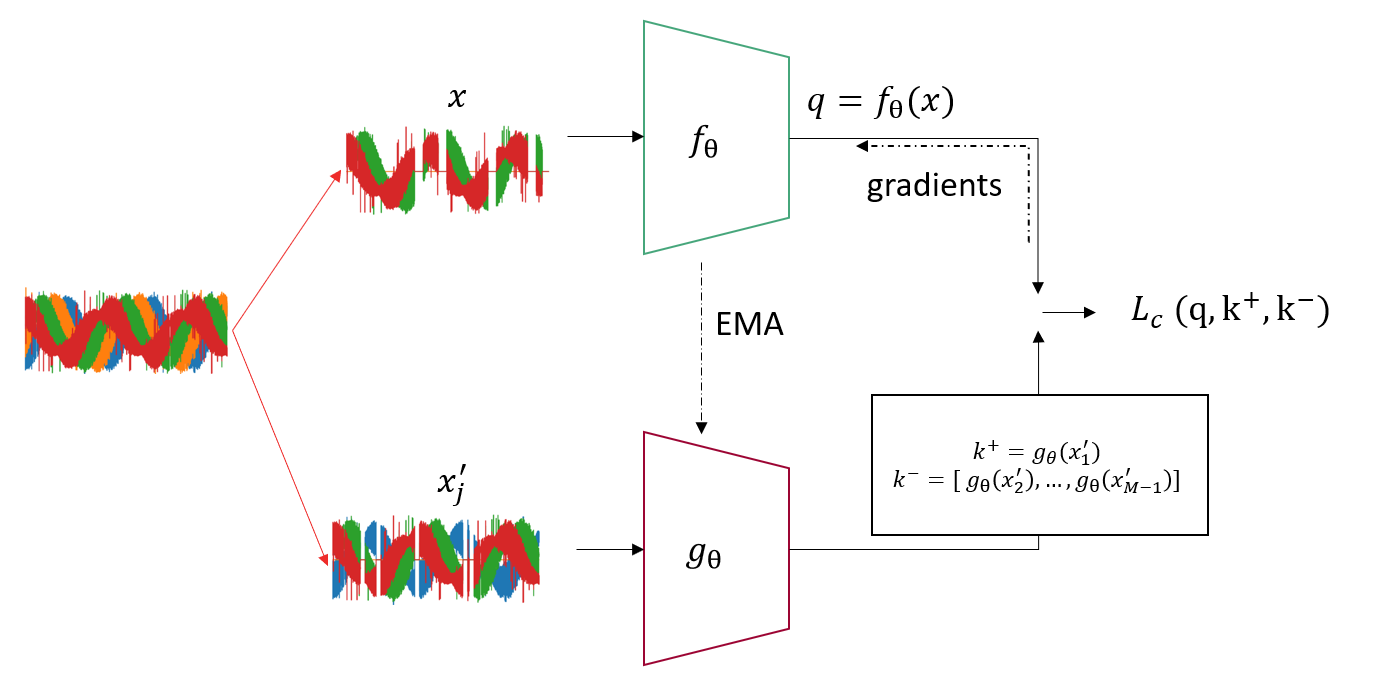}
        \caption{Self-supervised pretraining with MoCo framework.}
        \label{fig:moco-model}
    \end{subfigure}
    \begin{subfigure}[b]{0.45\linewidth}
        \centering
        \includegraphics[width=0.95\linewidth]{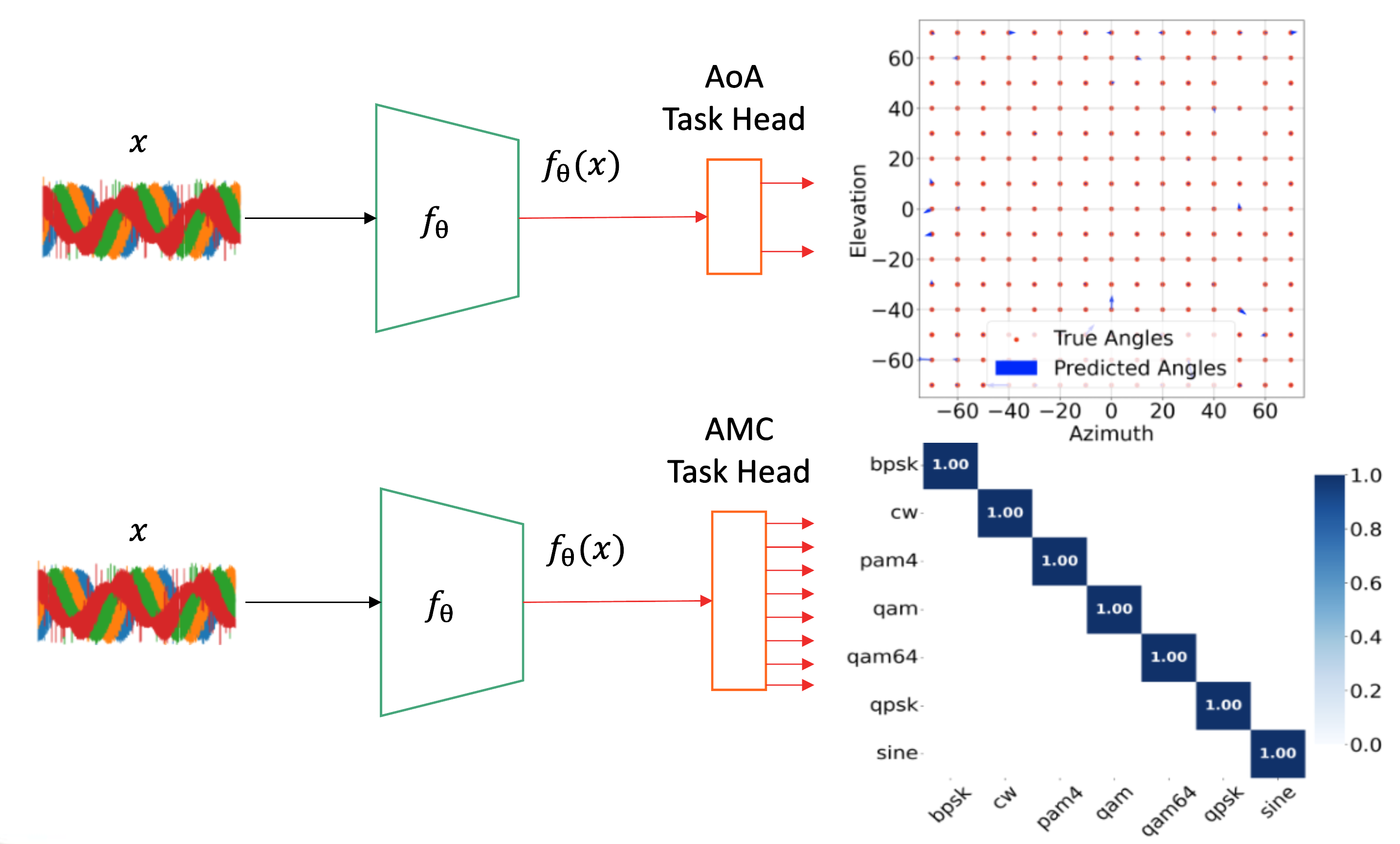}
        \caption{Downstream task fine-tuning for AoA and AMC tasks.}
        \label{fig:fine-tune}
    \end{subfigure}
    \caption{Proposed Self-Supervised Radio Representation Learning for Multitask Applications.}
\end{figure*}

\subsection{Problem Statement}

We consider a MIMO setup with $N_t$ transmit antennas and  $N_r$ receive antennas as show in \ref{fig:testbed}. The transmitter sends a variety of modulation signals and the receiver array captures the transmitted signals. The received signals are perturbed by propagation effects such as multipath and noise, at angles relative to the transmitter. The angles are the angle of elevation, $\theta$ and the azimuth angle, $\phi$, which are controlled by a servo motor at the receiver. The dataset is also generated with a set of different modulations, $m$. These are 16-QAM, 64-QAM, BPSK, QPSK, PAM4, and continuous wave (CW) AM modulation.

We therefore denote a sample from our dataset  $D$ as $\lbrace \mathcal{X}_s, (\theta, \phi), $ m$\rbrace^S_{s=1}$ where $\mathcal{X}_s$ is the set of $S$ IQ samples received at the $N_r$ antennas corresponding to the angles $(\theta, \phi)$, and at a transmit modulation $m$. 
Our goal is to learn transferable radio representation for two tasks--predicting the angles, $(\theta, \phi)$ and the modulation class $m$ from the received sample $\mathcal{X}_s$.
In other words, we aim to devise a self-supervised learning approach where the model learns useful representation via
pre-training on the dataset, $D$, while discarding the labels to extract meaningful latent invariant representations from the signal data $\mathcal{X}$. Thereafter the model is fine-tuned on a subset of the dataset, $D_{train}$ containing labels.
In what follows, we present the solution that learns these transferable representations for the related tasks and alleviates the need for a massive labeled dataset.

\subsection{Self-supervised Learning with Momentum Contrast}
We train the self-supervised model using momentum contrast (MoCo) \cite{chen2021empirical}, as shown in Figure  \ref{fig:moco-model}.
In the field of self-supervised learning, several contrastive methods have emerged, with some gaining traction in the wireless communication domain. One notable framework is MoCo, which has evolved significantly since its introduction in 2020, achieving state-of-the-art performance in the vision domain. Motivated by its success, we investigate MoCo's potential for developing multitask few-shot and zero-shot learners.
The MoCov3 framework has two branches of the encoder networks used to translate the input signal to a low-dimensional representation. The objective of the network is to pull the two low-dimensional representations of different augmented inputs (views), of an instance close together and away from all other instances in the dataset. Concretely, let $x$ and $x_j'$ be two augmented views of the network. The views are encoded by two neural networks, $f_\theta$ and $g_\theta$ such that $q = f_\theta(x)$ and $k_j = g_\theta(x'_j)$, assuming they are parameterized by weights, $\theta$. One network is trained while the other is slowly updated with an exponential moving average (EMA). This is done to avoid learning a trivial encoding.

MoCo aims to minimize the contrastive loss function and maximize invariant latent representations that are transferrable to downstream (related) tasks. 
The contrastive loss function \cite{chen2021empirical} is given as follows
\begin{align}
    \label{eqn:con_loss}
    \mathfrak{L}_{c} = \frac{1}{B}\sum_{i=1}^{B}\sum_{j=1}^{M}CrossEntropyLoss\left(\frac{q_i \cdot k_j}{\tau}, y_i \right) \cdot 2\tau.
\end{align}

The loss function is computed across a batch size $B$ and memory bank $M$ to both maximize and minimize the dot product between the query and key $q_i\cdot k_j$ for a pair of inputs (views). The positives and negatives in a batch are determined by the labels $y_i$. Intuitively, we compute the vector similarity between the stored representations, keys, and the updated (online) encoder, query. Lastly, the temperature parameter $\tau$ is a hyper-parameter that scales the loss.

For the i-th example in a batch, B, the contrastive loss function in Equation \ref{eqn:con_loss}, has a pair of positives, $k^+$, and M - 1 negatives, $k^-$. With augmentations, the network learns encodings such that the dot product $q_i\cdot k_j$ is maximized for different augmented views of the same example ($i=j$) and minimizes the dot product with all other keys. The algorithm is presented in Algorithm \ref{alg:moco-algorithm}.


\begin{algorithm}[!b]
\caption{MoCo-v3 Algorithm for Self-supervised Radio Representation Learning with IQ Data}
\label{alg:moco-algorithm}
\begin{algorithmic}[1]
\Procedure{MoCo-v3}{\textbf{x}}
\State \textbf{Input:} \parbox[t]{\dimexpr\linewidth-\algorithmicindent}{%
                                    Momentum encoder $g_\theta$, \\
                                    query encoder $f_\theta$,\\
                                    RF augmentation function $T_{IQ}(\cdot)$, \\
                                    and momentum update coefficient $m$
                                    }
\For{$\mathbf{x} \in \mathcal{X}$} \Comment{\parbox[t]{0.4\linewidth}{%
Load a batch of $B$ IQ data samples%
}}
    \State $\mathbf{x_1} = T_{IQ}(\mathbf{x})$ and $\mathbf{x_2} = T_{IQ}(\mathbf{x})$ 
    \State $\mathbf{q_1} = f_\theta(\mathbf{x_1})$ and $\mathbf{q_2} = f_\theta(\mathbf{x_2})$ 
    \State $\mathbf{k_1} = g_\theta(\mathbf{x_1})$ and $\mathbf{k_2} = g_\theta(\mathbf{x_2})$ 
    \State \parbox[t]{0.9\linewidth}{%
    Compute the symmetric contrastive loss in Eqn. (\ref{eqn:con_loss}).%
    }
    \State Backpropagate the loss.
    \State \parbox[t]{0.7\linewidth}{%
    Update the query encoder parameters $f_\theta$ using gradients.%
    }
    \State $g_\theta \leftarrow m \cdot g_\theta + (1 - m) \cdot f_\theta$
    \Comment{\parbox[t]{0.2\linewidth}{%
    Update the momentum encoder parameters%
    }}
\EndFor
\EndProcedure
\end{algorithmic}
\end{algorithm}

\subsubsection{Data Augmentation Methods}
\begin{figure}[h]
    \centering
    \includegraphics[width=.99\linewidth]{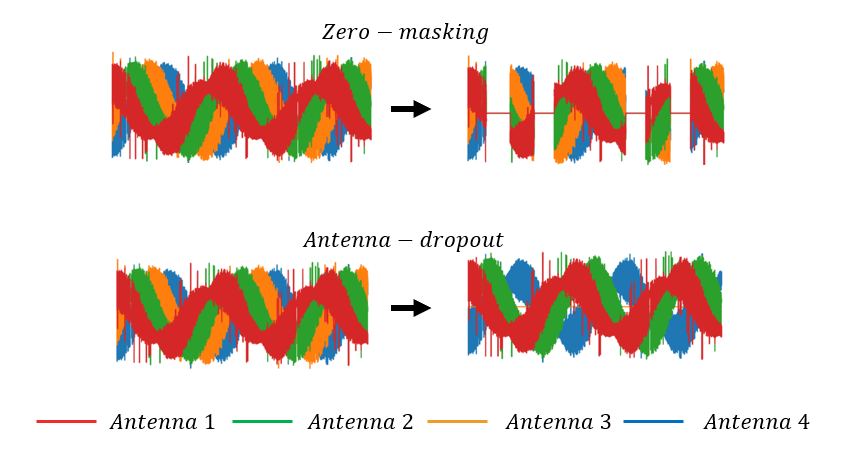}
    \caption{Augmentations for wireless contrastive learning. The original data shown on the left is transformed to the right after applying the augmentations of zero masking and antenna dropout.}
    \label{fig:augs}
\end{figure}

In the radio domain, several augmentations are possible, including DC shift, amplitude scaling, zero-masking, additive white Gaussian noise (AWGN), antenna dropout, and phase noise. These transformations alter numerical values but retain the semantic structure of the data, such as constellation regions and frame structure. For this task, we selected two transformations that demonstrated their effectiveness, with plans to investigate additional augmentations in future work. Figure \ref{fig:augs} illustrates the augmentations applied on the original signal, $X$. The x-axis denotes the time while the y-axis denotes the amplitude of the signal. The transformations are summarized in Table \ref{tab:augmentations}. In our experiments we set $p _1 = 0.2$ and $p_2 = 0.2$.

\begin{table}[!h]
    \centering
    \caption{Augmentation transformations}
    \label{tab:augmentations}
    \resizebox{\columnwidth}{!}{
    \begin{tabular}{c|l}
         \textbf{Augmentation} & \textbf{Transformation}  \\
         \hline
         Antenna Dropout\cite{hao2024bootstrapping} & $X' = \sum_j b_jX, b_j \sim$ Bernoulli($p_1$), for $0<j\leq N_r$\\
         Zero Masking\cite{davaslioglu2022self} & $X' = \sum_k m_k^{block}X, m_k^{block} \sim$ Bernoulli($p_2$) \\
    \end{tabular}
    }
\end{table}


\subsection{Fine-tuning}
The end goal of contrastive learning is to obtain a good $f_\theta$ that can used in a variety of wireless applications. A task-specific model is then constructed using the output of $f_\theta$ as features combined with a classifier head as shown in figure \ref{fig:fine-tune}. Typically, the wireless tasks also fine-tune (i.e., bias) these pre-trained representations towards application-specific criteria.

\subsection{Implementation Details}
The model architectures used for this study are MobileNetV3, ResNet18 and ResNet50.
The dataset (as discussed in section~\ref{ss:testbed_n_dataset}) is 
preprocesed to a 4-channel tensors with real and imaginary channels, $X' \in \mathcal{R}^{4\times2\times1024}$, where 1024 is the number of receive samples per training example. The deep learning network uses the MoCov3 \cite{chen2021empirical} framework to produce a 256-dimensional feature vector, such that $q,k \in \mathbb{R}^{1\times256}$ used in the contrastive loss.

\section{Experimental Setup}
\subsection{Testbed \& Dataset}\label{ss:testbed_n_dataset}

The dataset used in this study was generated using a software-defined radio (SDR) testbed. It contains IQ wireless signal data captured outdoors using a USRP X300 transmitter and two synchronized USRP X300 receivers using an Ettus Octoclock CDA-2990. Each of the X300 USRPs have 2 Tx and 2 Rx channels and they were equipped with 5.88 GHz patch antennas. There is a total of 4 Rx receiver antennas and the IQ signals at different AoA were captured across the 4 channels.
All I/Q data was recorded for post-processing, ensuring synchronized, high-quality data capture essential for accurate angle of arrival (AoA) detection across multiple channels. The dataset used for this study was generated with the testbed previously shown in Figure \ref{fig:testbed}.



Each recording is saved as a numpy (.npy) file, containing 4x1000000 complex I/Q samples, alongside metadata specifying the azimuth and elevation angles for each recording. The dataset covers various positions and angles, with azimuth and elevation ranging from -70\textdegree\ to 70\textdegree, controlled precisely by servo motors.
The elevation servo is fixed to the ground, with the azimuth servo mounted on top, allowing for precise, repeatable adjustments of the receiver’s orientation relative to the transmitter. By rotating the receiver antenna, this configuration effectively simulates different transmitter locations within the receiver’s reference frame, offering a comprehensive range of AoA detection scenarios.

The total recordings captured is 68 GB constituting 4,609 number of files, the dataset was split into 3,687 files for the development set and 922 files for the test set. This split before pre-processing avoids data leakage and the pre-processing step slices each recording into a chunk of 1024 sample length.
The numpy recordings are balanced across angles and modulations with approximately 20 recordings per unique angles and 658 recordings per modulation. The development set has 1,784,508 examples and the test set has 446,248 examples.


\subsection{Training}
The training is an adaptation of the MocoV3 framework \cite{chen2021empirical} which uses layer-wise adaptive rate scaling (LARS) 
optimizer, a learning rate of 0.03, weight decay of $10^{-4}$, momentum set to 0.996 and a batch size of 1024. The learning rate is adjusted with the batch size with \textit{$\text{learning rate} \times \text{batch size} / 256$} similar to \cite{chen2021empirical} which decays according to a cosine schedule. Also, the momentum is adjusted during training gradually to 1 using the cosine function. We train for approximately $3 - 6$ hours. In addition, training was done with mixed precision training to speed up training time and to reduce the size of training parameters needed to fit the GPU resources. We trained the MobileNetv3 and ResNet18 models with one T4 GPU and the ResNet50 model with two T4 GPUs on Google Cloud resources.

For the fully supervised baseline model and fine-tuning of the linear classifier head in the downstream tasks, we train with AdamW or SGD optimizers using a learning rate of $10^{-2}$  or $10^{-3}$ and a weight decay of $10^{-2}$ when applicable. We train for 50 epochs for both steps. Table ~\ref{tab:training_params} summarizes the details of the training parameters both for pretraining and finetuning.
\begin{table}[!h]
    \centering
    \caption{Radio Representation Learning Training Parameters}
    \label{tab:training_params}
    \resizebox{\columnwidth}{!}{
    
    \begin{tabular}{c|c|c}
        \textbf{Parameters} & \textbf{Pre-training} & \textbf{Fine-tuning} \\
        \hline
        Optimizer & LARS & AdamW or SGD \\
        \hline
        Learning rate & $0.3 \times \text{batch size} / 256$ & 0.01 or 0.001 \\
        \hline
        Weight Decay & $10^{-4}$ & $10^{-2}$ if applicable \\
        \hline
        Batch Size & 1024 & 1024 or 512 \\
        \hline
        Epochs & 20 ($\sim 3 - 6$ hours) & 50 ($\sim 7 - 20$ hours) \\
        \hline
        MoCo Temperature & Dynamic between 0.1 and 0.2 \cite{manna2023dynamically} & - \\
        \hline
        MoCo Momentum & 0.996 & - \\
        \hline
        MoCo Momentum Adjustment & Cosine & - \\
        \hline
        MoCo MLP dimension & 1024 & - \\
        \hline
        MoCo Output dimension & 256 & - \\
        \hline
        LR Scheduler & Cosine Annealing & Cosine Annealing \\
    \end{tabular}
    }
\end{table}

\subsection{Performance Metrics \& Tasks}

\subsubsection{Downstream Tasks}
In this study, we consider two downstream tasks: angle of arrival (AoA) estimation and Automatic Modulation Classification (AMC).

\subsubsection{Metrics}
We evaluate our methods by the performance of the encoder $f_\theta$ on the downstream tasks with frozen and fine-tuned weights. For the angle of arrival task, the metric of evaluation is the mean absolute error (MAE) across all angles in the evaluation dataset, while the metrics of evaluation for the automatic modulation classification are accuracy, precision, and recall. The use of MAE is for ease of interpretation, another similar metric used in literature is the root mean square error (RMSE). These performances are compared with those of the fully supervised counterparts.

\subsubsection{Data Efficiency}
Additionally, we investigate the data efficiency of the contrastive representation $f_\theta$. Data efficiency refers to the amount of labels required to build a downstream task using $f_\theta$. This is evaluated for both tasks and
the metrics are evaluated against their fully supervised counterparts.

\section{Results \& Analysis}
In this section, we present a thorough analysis of the results we achieved.
\subsection{Performance on Downstream Tasks}

We present the results of our best-performing pre-trained model, ResNet50, on two downstream tasks; angle of arrival (AoA) and automatic modulation classification (AMC). We conduct these experiments with two different settings of the encoder weights discussed below.

\subsubsection{Frozen Encoder Weights}
 Frozen weights means that the learned representations/models of the encoder are not updated during training. Having a good performance with frozen weights is challenging in representation learning, and indicates that the learned representations are effective and generalizable. 

\begin{table}[!b]
    \centering
    \caption{Evaluation of pre-trained model on different RF tasks with \emph{frozen} feature extractor weights compared to randomly initialized network.}
    \label{tab:frozen}
    \resizebox{\columnwidth}{!}{
    \begin{tabular}{c|c|c c c}
        Tasks & \textbf{AoA} & \multicolumn{3}{c}{\textbf{AMC}} \\
        \hline
        Metrics & MAE (\textdegree) $\downarrow$ & Acc. (\%) $\uparrow$ & P (\%) $\uparrow$ & R (\%) $\uparrow$ \\
        \hline
        Xavier init. & 8.93 & 92.16 & 92.59 & 92.36 \\
        Ours & \textbf{4.28} & \textbf{99.38} & \textbf{99.40} & \textbf{99.39} \\
        \textbf{IMPROVEMENTS} & \textcolor{blue}{+52.1\%} & \textcolor{blue}{+7.83\%} & \textcolor{blue}{+7.35\%} & \textcolor{blue}{+7.61\%} \\
    \end{tabular}
    }
\end{table}

\begin{table}[!b]
    \centering
    \caption{Evaluation of pre-trained model on different RF tasks with \emph{fine-tuned} feature extractor weights compared to fully supervised network.}
    \label{tab:fine-tuned}
    \resizebox{\columnwidth}{!}{
    \begin{tabular}{c|c|c c c}
        Tasks & \textbf{AoA} & \multicolumn{3}{c}{\textbf{AMC}} \\
        \hline
        Metrics & MAE (\textdegree) $\downarrow$ & Acc. (\%) $\uparrow$ & P (\%) $\uparrow$ & R (\%) $\uparrow$ \\
        \hline
        Supervised & 0.73 & 99.811 & 99.817 & 99.812 \\
        Ours & \textbf{0.71} & \textbf{99.985} & \textbf{99.982} & \textbf{99.985} \\
        \textbf{IMPROVEMENTS} & \textcolor{blue}{+0.03\%} & \textcolor{blue}{$\sim$0.00\%} & \textcolor{blue}{$\sim$0.00\%} & \textcolor{blue}{$\sim$0.00\%} \\
    \end{tabular}
    }
\end{table}

We evaluate the frozen weights against randomly initialized weights of the feature extractor and the results for the two downstream tasks, AoA and AMC are presented in Table \ref{tab:frozen}. This is evaluated for 15 epochs of training using two linear classifier heads for the AoA task and one linear classifier head for AMC attached to the encoder, $f_\theta$.
We achieve up to 52.1\% improvements over random initialization for the more challenging AoA task, bringing down the MAE to a more reasonable 4 degrees of error. 
For the simpler AMC task, our self-supervised representation learning approach has been able to achieve close to 100\% accuracy.


\subsubsection{Fine-tuned Encoder Weights}
With fine-tuned weights, the contrastive learning approach allows the encoder weights to be fine-tuned for the each tasks during downstream task training. For good representations, the performance in this scenario is expected to be close to the trained from scratch supervised solution \cite{chen2021empirical} and our results presented in Table \ref{tab:fine-tuned} demonstrate that.
To achieve the results, we fine-tuned the encoder model with a learning rate in the order of $10^{-1}$ compared to the linear classifier layer. Additionally, we observed that a linear warm-up period is best suited when fine-tuning the pre-trained weights. The results show competitive performance between the fully trained model and the fine-tuned contrastive model.

\subsection{Impact of More Classifier Layers}

\begin{figure}[!t]
    \centering
    \begin{subfigure}[b]{0.45\linewidth}
        \centering
        \includegraphics[width=\linewidth]{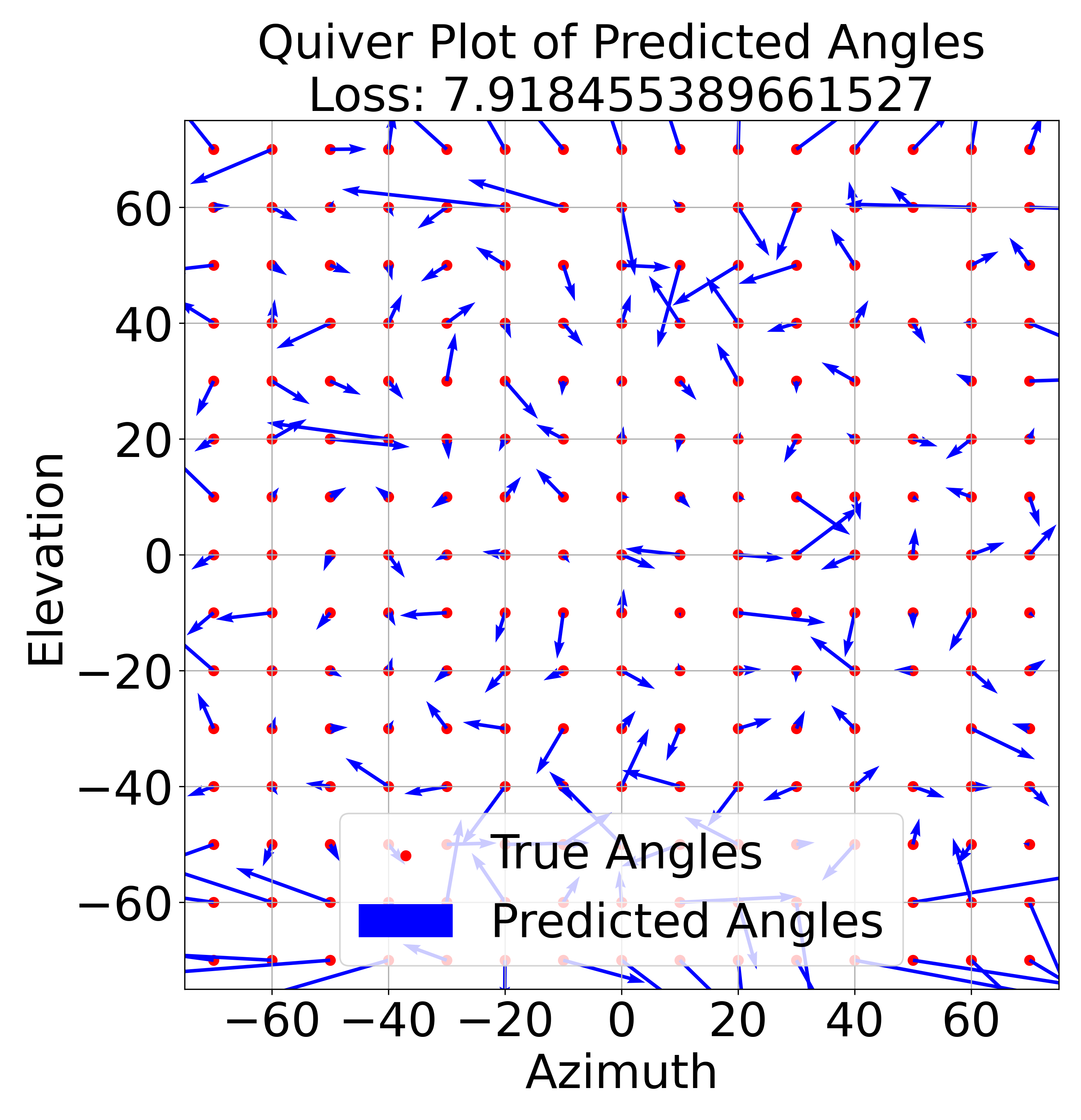}
        \caption{One classifier layer}
        \label{fig:aoa-frozen_1x}
    \end{subfigure}
    \begin{subfigure}[b]{0.45\linewidth}
        \centering
        \includegraphics[width=\linewidth]{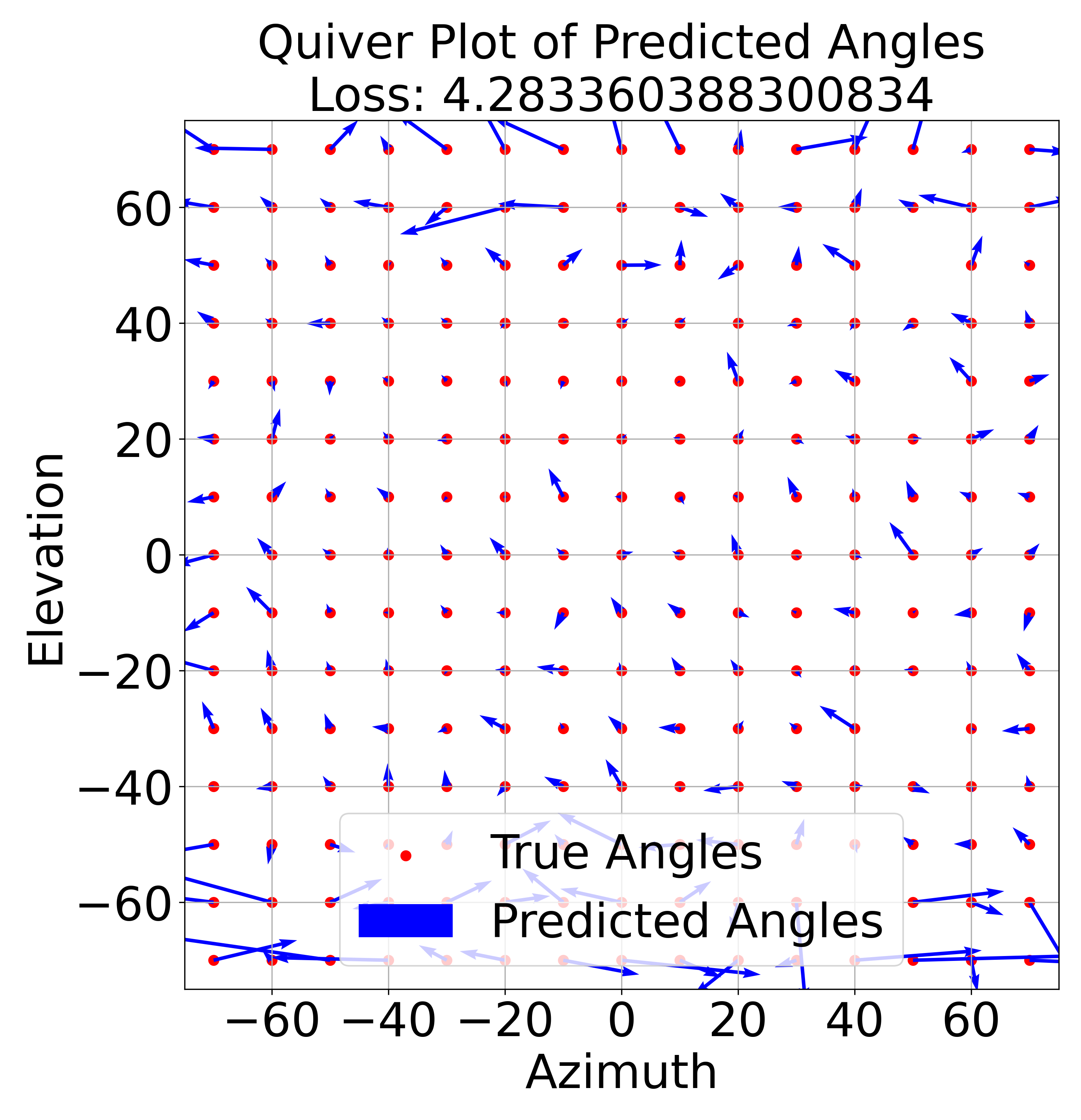}
        \caption{Two classifier layers}
        \label{fig:aoa-fine_tuned_2x}
    \end{subfigure}
    \caption{Quiver plots for all angles in the angle of arrival estimation task, represented by red markers and the mean error of all predictions, represented as blue arrows. Visual demonstration of improved performance with \textit{frozen} encoder weights when the classifier layer is \textit{doubled}.}
    \label{fig:2x-vs-supervised}
\end{figure}

With frozen encoder weights, the performance on the AMC task was substantially high but the performance on the AoA was initially 7.92 degrees of error. In an attempt to improve the performance with frozen encoder weights, we investigated the impact of more linear classifier layers. This would allow more complex non-linear relationships to be made from the encoder embeddings.
By doubling the number of linear layers, we improve the performance by 46\% from 7.92 to 4.28 degrees of error. Figure \ref{fig:2x-vs-supervised} shows a visual comparison of the performance of the fine-tuned linear classifier with two layers versus one layer as quiver plots. There is clear indication that the performance is significantly better with only some of the larger angles exhibiting errors.

\subsection{Impact of Model Size}

\begin{figure}[!t]
     \centering
     \includegraphics[width=0.85\linewidth]{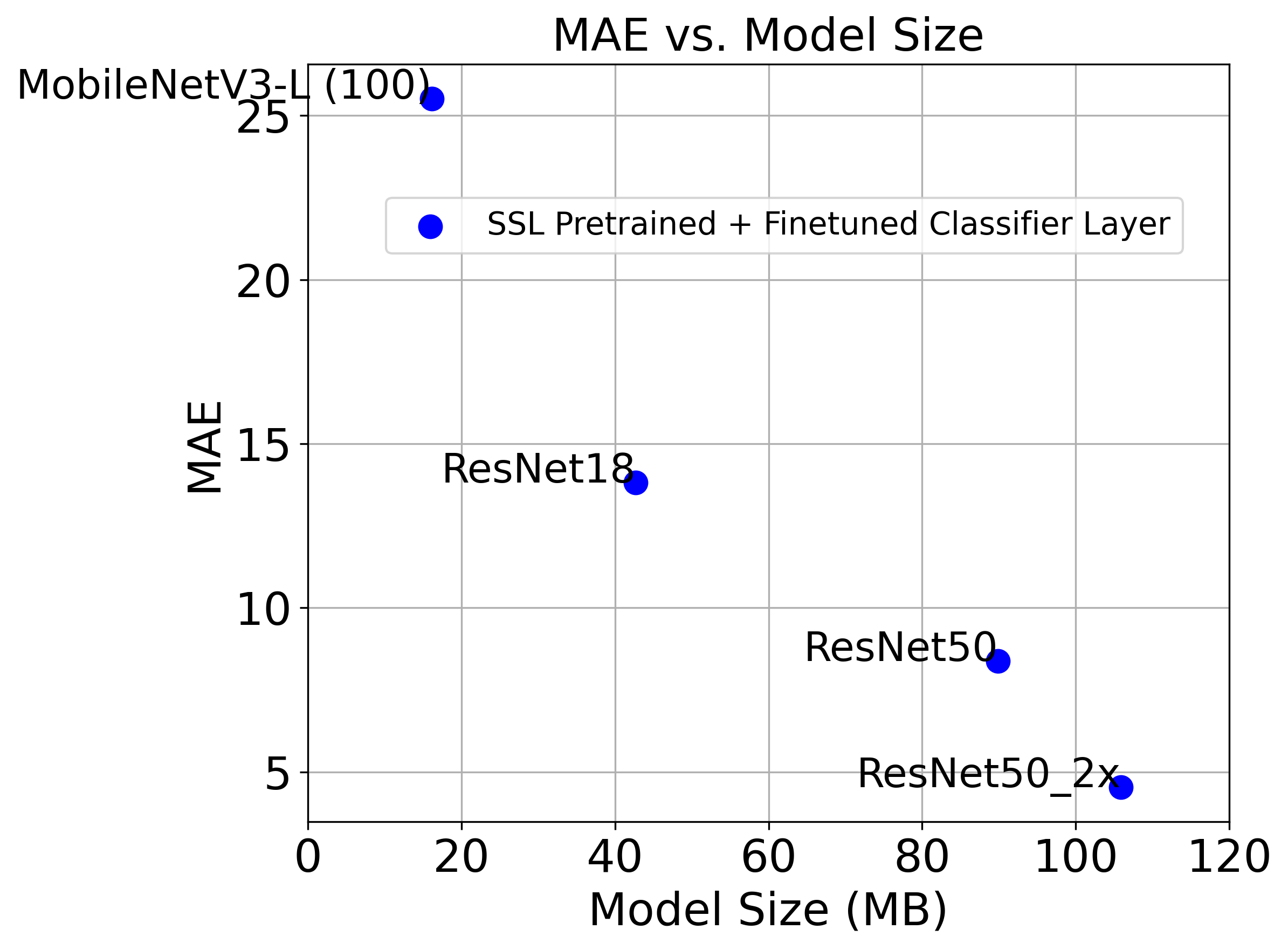}
     \caption{Model scaling effectively reduces MAE differences between SSL pretrained and supervised learning approaches.}
     \label{fig:model-size}
 \end{figure}

In this study, we explored different model sizes and architectures to evaluate the effects on the performance of the proposed SSL radio representation learning approach.
We observed that with larger models such as ResNet50, performance improves compare to smaller models such as the ResNet18 and MobileNet architectures.  
This trade-off is summarized in Figure \ref{fig:model-size} which shows 
that with more learning capacity, the MoCo pretraining is more effective. In addition, to the impact of the different models, the impact of increasing the linear classifier layer is also shown in the Figure, indicated by ResNet50\_2x. This further improves the performance of frozen encoder weights pretrained with SSL. 

\subsection{Data Efficiency}

We investigated the data efficiency of the learned representation against the supervised counterpart using a ResNet50 model trained on four data ratios, $D_{train}$--0.1\%, 1\%, 10\% and 100\%. As discussed in section \ref{ss:testbed_n_dataset}, we have a dataset of 69 GB of size which has a development set of 1,784,508 examples and a test set of 446,248 examples. We applied the data ratios on the development set such that the lowest ratio of 0.1\% results in a dataset of size 1,784 examples and present our results based on the performance on the test set. Table \ref{tab:data-eff-combined} shows the results of this study which is produced using the same training configurations introduced earlier. 
In the AoA task, we fine-tuned the encoder weights with a very small learning rate in the order of $10^{-2}$ compared to the classifier layer. In all other data ratios, the pre-trained weights are trained end-to-end. In the AMC task, we achieved better results when we preserved the pretrained weights of the encoder as much as possible, thus in the 0.1\% data ratio, we used a learning rate in the order of $10^{-3}$ compared to the classifier layer and the order of $10$ compared to the classifier layer for all other data ratios.

The proposed method shows strong data efficiency across both tasks, especially at low data regimes. The fully supervised counterpart has similar performance only at full data scale, where a 100\% of the labeled data is used. This shows that the proposed SSL approach was indeed significantly better at leveraging minimal labeled data.

\subsection{Ablation Study of Augmentations}

The choice of augmentations matters in contrastive learning. 
To optimize the pre-training stage, we examined augmentations of IQ data using an ablation study. Effective augmentations should support the contrastive objective by promoting invariance, maximizing similarity across positive pairs, and repulsion across negatives. In a noise-free dataset, the unaugmented data serves as a good baseline. 
Table \ref{tab:aug-ablation} shows the augmentations and the results compared to the baseline. This is achieved with frozen pretrained encoder weights on the AoA task.
For this analysis, we kept all hyperparameters as indicated in Table \ref{tab:augmentations}, $p_1$ and $p_2$ at 0.2. As show, there is a significant impact of augmentation on the learned representation, as there is a significant margin compared to the baselines.

\begin{table}[!t]
    \centering
    \caption{Data Efficiency measurement for the AoA and AMC tasks. 
    Smaller MAE values indicate better performance for AoA, while higher accuracy values indicate better performance for AMC.}
    \label{tab:data-eff-combined}
    \begin{tabular}{c|cccc}
         & \multicolumn{4}{c}{\textbf{AoA Task (MAE) $\downarrow$}} \\
        \hline
         Method & 0.1\% & 1\% & 10\% & 100\% \\
        \hline
        Supervised & 18.90 & 3.85 & 1.68 & 0.73 \\
        Ours & \textbf{10.98} & \textbf{3.51} & \textbf{1.45} & \textbf{0.71} \\
        \textbf{IMPROVEMENTS} & \textcolor{blue}{+41.90\%} & \textcolor{blue}{+8.83\%} & \textcolor{blue}{+13.69\%} & \textcolor{blue}{+0.027\%} \\
        \hline
        \hline
         & \multicolumn{4}{c}{\textbf{AMC Task (Accuracy) $\uparrow$}} \\
        \hline
         Method & 0.1\% & 1\% & 10\% & 100\% \\
        \hline
        Supervised & 39.190 & \textbf{98.030} & 98.907 & 99.811 \\
        Ours & \textbf{48.863} & 97.951 & \textbf{99.169} & \textbf{99.985} \\
        \textbf{IMPROVEMENTS} & \textcolor{blue}{+24.68\%} & \textcolor{red}{-0.0008\%} & \textcolor{blue}{+0.003\%} & \textcolor{blue}{+0.002\%} 
    \end{tabular}
\end{table}



\begin{table}[!h]
    \centering
    \caption{Augmentation Ablation study. Impact of one augmentation at a time on AoA evaluated with frozen encoder weight and classifier layer fine-tuning.}
    \label{tab:aug-ablation}
    \resizebox{0.75\columnwidth}{!}{
    \begin{tabular}{c|c}
        Augmentations & AoA $\downarrow$ (MAE) \\
        \hline
         Antenna Dropout \cite{hao2024bootstrapping} & 8.94 \\
         Zero Masking \cite{davaslioglu2022self} & \textbf{7.43} \\
         \hline
         No Augmentation (baseline) & 30.43 \\
         \hline
    \end{tabular}
    }
\end{table}

\section{Conclusion}
This work has proposed and successfully demonstrated the applicability of momentum contrastive self-supervised learning in the wireless radio domain for two tasks: angle of arrival (AoA) estimation and automatic modulation classification (AMC). Despite known challenges in applying contrastive learning to radio signals, our results indicate that with carefully designed augmentations and diverse data, contrastive learning can yield quality, invariant latent representations. We anticipate that incorporating additional augmentations, and more diverse datasets will further enhance performance and applicability to even more than two tasks. 
In summary, our findings highlight the potential of self-supervised learning to transform wireless communication tasks by reducing dependence on labeled data and improving model generalization, paving the way for scalable foundational 6G AI-native models. 



\section*{Acknowledgment}


We appreciate Dr Mohammed Alloulah for his insightful discussions on self-supervised contrastive learning and radar MIMO augmentation. Those discussions sparked ideas for the work and contributions the paper presents.

\bibliographystyle{ieeetr}
\bibliography{References}

\end{document}